\documentclass{elsart}

\usepackage{natbib}

\usepackage{graphicx}

\usepackage{amssymb,amsmath}

\begin{document}

\begin{frontmatter}


\title{Gravitation Physics at BGPL}
\author{P. E. Boynton\corauthref{cor1},}
\ead{boynton@phys.washington.edu}
\corauth[cor1]{}
\author{R. M. Bonicalzi,}
\author{A. M. Kalet,}
\author{A. M. Kleczewski,}
\author{J. K. Lingwood,}
\author{K. J. McKenney,}
\author{M. W. Moore,}
\author{J. H. Steffen}
\address{
University of Washington\\
MS 351560\\
Seattle, WA 98195, USA
}

\author{E. C. Berg,}
\author{W. D. Cross,}
\author{R. D. Newman}
\address{
University of California at Irvine\\
4129 Frederick Reines Hall\\
Irvine, CA 92697, USA
}

\author{R. E. Gephart}
\address{
Pacific Northwest National Laboratory\\ MSIN K8-88 Richland, WA\\
99354, USA }

\begin{abstract}
We report progress on a program of gravitational physics experiments using cryogenic torsion pendula undergoing large-amplitude torsion oscillation.  This program includes tests of the gravitational inverse square law and of the weak equivalence principle.  Here we describe our ongoing search for inverse-square-law violation at a strength down to $10^{-5}$ of standard gravity.  The low-vibration environment provided by the Battelle Gravitation Physics Laboratory (BGPL) is uniquely suited to this study.
\end{abstract}


\end{frontmatter}

\section{Introduction}
The torsion pendulum remains the instrument of choice for a
variety of laboratory tests of gravitational phenomena.  Such
experiments necessarily require the detection of extremely small
torques.  First employed in the 18$^{\rm th}$ century, this
instrument was operated in a mode in which the signal torque was
detected as an angular deflection of a pendulum in the presence of
a source mass.  This ``deflection method" is susceptible to
certain systematic errors, even a small tilt of the apparatus.
One way to overcome this drawback is to operate with the pendulum
executing torsional oscillations and to detect the presence of a
small torque on the pendulum arising from interaction with the
source mass through the measured shift in oscillation frequency
correlated with changes in relative orientation between the
pendulum equilibrium position and the direction to the source
mass\citep{uwmoriond}.

This ``frequency method" was commonly used in the ${\rm 20^{th}}$
century for measurements of the gravitational constant {\it G},
but with milliradian oscillation amplitude.  We operate at a much
larger amplitude, near the value that yields a maximum in
signal-to-noise ratio\citep{uwa2}.  A downside to the frequency
method has been the temperature sensitivity of a torsion fiber's
elastic modulus.  Variation in fiber temperature will produce
systematic and random errors in the frequency measurement.  An
effective remedy is to operate the pendulum in a cryogenic
environment, providing reduced temperature sensitivity and an
opportunity for improved temperature control.

Details of our cryogenic torsion pendulum system, which uses this
frequency method, can be found
elsewhere\citep{mg7,moriond,london,uzbek}.  The cryogenic
environment has a number of benefits besides temperature
stability, including low thermal noise (${\it
\sim\sqrt{\frac{k_BT}{Q}}}$), high fiber {\it Q} ($>$ ${\rm
10^5}$), slower drift of angular equilibrium position, easily
achieved high vacuum, and effective magnetic shielding with
superconducting materials.  Concerns of bias due to fiber
nonlinearity and anelasticity have been addressed and appear not
to be significant\citep{icifuas}.

\begin{figure}[ht]
\begin{center}
\includegraphics[width=0.5\textwidth]{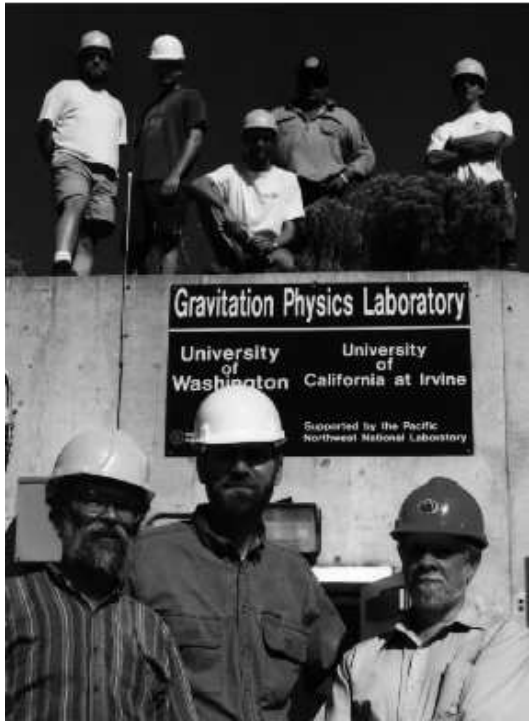}
\caption{Riley Newman (UCI), Roy Gephart (PNNL) and Paul Boynton (UW) with students and staff in backgound at the entrance to BGPL.\label{labpic}}
\end{center}
\end{figure}

Our apparatus is located at the Battelle Gravitational Physics Laboratory (BGPL) on the Department of Energy Hanford site in the arid lands of eastern Washington State.  This facility is several kilometers from public access and associated anthropogenic noise, and was developed largely because of its low ground-motion background (power density less than ${\rm 10^{-22}}$ ${\rm m^2/Hz}$ at 10 Hz).  The photo in Figure \ref{labpic} shows the entrance to this underground facility developed by the UW/UCI collaboration with the support of the Pacific Northwest National Laboratory.

The siting of BGPL in 1994 was determined by a two-year vibration
survey of a number of candidate locations in Washington State.  A
former Nike missile launch facility on the DOE-Hanford site was
chosen to house our laboratory not only because it was
significantly quieter than all other locations surveyed, but also
was found to exhibit an exceptionally quiet vibration environment
by any standard.  This attribute, which is essential to the
success of our program, follows from three circumstances:  1) A
low background of natural seismic motion because the exceptionally
rigid foundation of BGPL is fairly tightly coupled to the massive
basalt formations that extend east of the Cascade Range.  2) The
laboratory is situated underground with virtually no structure
above grade level; consequently, wind buffeting is minimal.  The
absence of trees is a significant advantage for the same reason.
3) There is virtually no anthropogenic contribution to ground
motion at this location due to highly restricted access to the
Hanford site.

This third item is the most important because human activity
commonly overwhelms the natural seismic background and fixes a
substantially higher level of ambient ground vibration than would
otherwise be present.  This low level of anthropogenic noise is
primarily responsible for the unique suitability of the BGPL for
carrying out extremely sensitive experiments on the frontier of
gravitation physics.  Equally important for BGPL is that
restricted access within the Hanford site perimeter (6 km distant)
assures this low background {\it remains protected} from common
anthropogenic vibration sources.  For this same reason, our
instrumentation is fully automated as data cannot be taken until
personnel vacate BGPL.

\section{Experimental Approach}
Along with tests of the equivalence principle, it is important to verify the weak-field limit of general relativity in a manner that is composition independent; namely, through tests of the inverse square force law.  An inverse square law violation (ISLV) could imply the existence of a superposed non-Newtonian force that couples approximately to mass.

Here we describe a null experiment in which a specially configured
torsion pendulum undergoing large-amplitude oscillations in
proximity to a source mass also of special form, is potentially
able to detect inverse square law violations approaching ${\rm
10^{-6}}$ of standard gravity at a range around ten centimeters
--- nearly a two-order-of-magnitude improvement over the current
empirical limit at that length scale\citep{gradiometer}.  This
improvement is accomplished by an experimental configuration that
is only second-order sensitive to fabrication errors in pendulum
and source mass, and not by a substantial reduction in dimensional
tolerances.

\begin{figure}[ht]
\begin{center}
\includegraphics[width=0.5\textwidth]{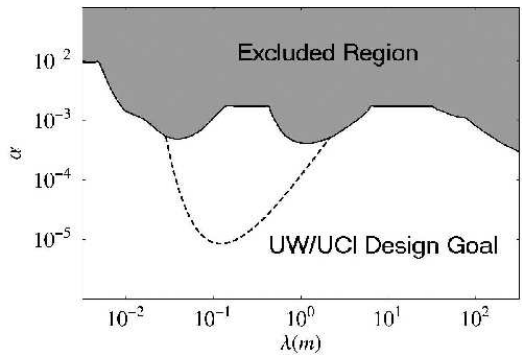}
\caption{Composition-independent, experimental, 2${\it \sigma}$ upper bound on ${\it \alpha}$ as a function of ${\it \lambda}$.  The dashed curve denotes the design goal of this ISLV experiment.
\label{alphalambda}}
\end{center}
\end{figure}

There are several ways to characterize a violation of the inverse square law.  The most common is to add a Yukawa term to the gravitational potential, yielding an interaction potential
\begin{eqnarray}
V&=&V_g+V_Y \nonumber \\[4pt]
{}&=&-\frac{GM}{r}\left(1+\alpha e^{-r/\lambda}\right)\, ,
\label{fullpot}
\end{eqnarray}
where {\it G} (at ${\it r=\infty}$) is Newton's constant, {\it M} the mass of the object generating the potential, ${\it \alpha}$  the strength of the ISLV potential relative to gravity, and ${\it \lambda}$  a characteristic length scale of the interaction.  Figure \ref{alphalambda} shows the current experimental constraint on ${\it \alpha}$ as a function of ${\it \lambda}$, and the projected measurement capability of our instrumentation.

While a Newtonian gravitational potential satisfies Laplace's
equation in a region with no gravitational sources, a
non-Newtonian (ISLV) interaction manifests itself as a {\it
nonzero} Laplacian of the potential in a source-free region.
However, because the Laplacian is a scalar and thus spherically
symmetric, it cannot produce a torque on our pendulum.  We
therefore designed our experiment to detect a nonzero horizontal
gradient of the Laplacian, the lowest multipole-order indication
of ISLV in a torsion experiment\citep{uwpakistan}.  Because we
directly detect the horizontal gradient of the Laplacian, our
experimental results will be independent of the ISLV model.

The total interaction energy between an object with density profile ${\it \rho(\textbf{{\it r}})}$ and the potential energy(\ref{fullpot}) is given by
\begin{equation}
U = \int \rho Vd^3r.
\label{energy}
\end{equation}

A Taylor expansion of the potential energy in cartesian coordinates about the origin yields
\begin{equation}
U = \int \rho \left( 1 + x\frac{\partial}{\partial x} + y\frac{\partial}{\partial y} + z\frac{\partial}{\partial z} + \frac{1}{2} x^2 \frac{\partial^2}{\partial x^2} + xy \frac{\partial^2}{\partial x \partial y} + \ldots \right)V dxdydz
\label{energyexp}
\end{equation}
where the derivatives are evaluated at the origin.  These terms can be grouped into expressions that have the symmetries of the spherical harmonics.  This gives a concise, orthonormal basis in which to express the interaction energy.  The general expansion of the interaction energy may then be written as
\begin{equation}
U = \sum_{nlm} V_{nlm} M_{nlm}
\end{equation}
where
\begin{equation}
M_{nlm} \propto \int \rho r^n Y_l^m d^3r
\end{equation}
and the ${\it V_{nlm}}$'s are the appropriate derivatives of the
potential evaluated at the origin.  Grouping the cartesian terms
of Equation (\ref{energyexp}) into the spherical harmonics is
nontrivial and normalization of the moments is also challenging.
This problem was examined by \citet{uwa1}.

The horizontal gradient of the Laplacian corresponds to the multipole term
\begin{eqnarray}
U_{311} &=& V_{311}M_{311} \nonumber \\[4pt]
{}&=& \left(\frac{1}{10}\right)\frac{\partial}{\partial x} \left( \nabla^2 V \right) \int \rho x \left(x^2+y^2+z^2 \right) dxdydz \nonumber \\[4pt]
{}& \propto &\frac{\partial}{\partial x} \left( \nabla^2 V \right) \int \rho r^3 Y_1^1 dxdydz\, .
\end{eqnarray}
We call ${\it M_{311}}$ and ${\it V_{311}}$ the 311 mass and field moments respectively.  The 311 mass moment is manifestly non-Newtonian because the field moment to which it couples, ${\it V_{311}}$, is a derivative of the Laplacian and would be identically zero if the interaction were strictly Newtonian.

There are two primary considerations that drove the pendulum and source mass design. First, to take advantage of the multipole expansion, the ratio of the size of the pendulum to the distance to the source mass is less than one, ensuring the contribution to the interaction energy from each higher order is suppressed approximately by that ratio.  Second, we purposely eliminated the low-order Newtonian mass and field moments of pendulum and source mass respectively, up to the order for which the contribution from the nonzero moments falls below estimated systematic errors.  In particular, since our signal has {\it m} = 1 symmetry, we null by design all {\it m} = 1 Newtonian pendulum mass moments from {\it l} = 1 to {\it l} = 6 and source-mass field moments from {\it l} = 1 to {\it l} = 8.

This latter constraint was suggested by previous experiments designed to detect ISLV using torsion pendulum techniques that relied on a pendulum with an exaggerated Newtonian moment.  Indeed, a high sensitivity to ISLV could be achieved using a barbell with large {\it l} = 2, {\it m} = 2 mass moment.  On the other hand, this exaggerated Newtonian mass moment would couple to the residual {\it l} = 2, {\it m} = 2 field moments (arising from  fabrication errors associated with the source mass) to produce a correspondingly large {\it m} = 2 systematic effect that would mimic a putative ISLV signal in such an experiment.  That is, any such scheme to increase the ISLV signal moment would necessarily increase the limiting systematic effect as well.

By designing a pendulum that has no low-order Newtonian moments we can reduce such systematic effects by two orders of magnitude.  Residual Newtonian moments resulting from fabrication errors in the source mass would not directly interact with the pendulum since it is specifically designed not to detect them.  It is therefore only the residual Newtonian mass moments resulting from errors in fabricating the pendulum that can couple to those residual Newtonian field moments.  As a result, departures from the design configuration are manifest only in second order in these errors.  This design strategy dramatically reduces systematic effects arising from standard machining tolerances, metrology, and material density inhomogeneity without resorting to heroic fabrication techniques.

\section{Experimental Design}

\begin{figure}[ht]
\begin{center}
\includegraphics[width=0.9\textwidth]{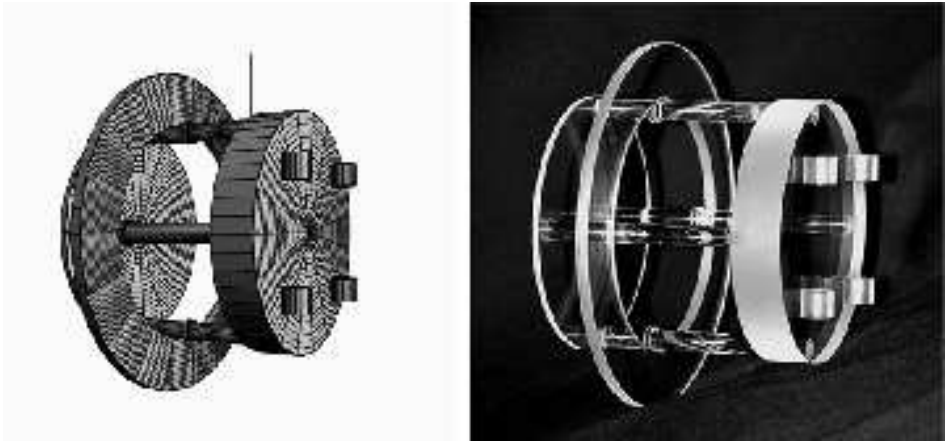}
\caption{Rendition of the ISLV pendulum on left and realization on right (prior to coating with gold).
\label{islvpend}}
\end{center}
\end{figure}

The result of pursuing this approach to pendulum design is shown
in Figure \ref{islvpend}.  This configuration was fabricated at UC
Irvine from fused silica parts.  It has a mass of 240 g and is
approximately 8 cm in diameter.  The pendulum is suspended by a
BeCu fiber that, under the weight of the pendulum, is stressed to
about 60\% of its tensile strength.  The pendulum is gold coated
and the fiber electrically grounded to suppress electrostatic
torques (important because the minimum detectable signal energy
for {\it S/N} = 1 is only on the order of tens of eV).  The
fabrication tolerance of the pendulum parts and their assembly is
6 ${\rm \mu}$m for most dimensions.  The parts are bonded together
using a room-temperature hydroxide catalysis technique that
locally dissolves the surfaces of the fused silica parts, fusing
them together in a siloxane bond and liberating water in the
process\citep{gwo}.  These bonds can withstand repeated cooling to
$\sim$2K when placed in the cryostat.  This technique facilitates
precise component positioning with minimal added mass.

\begin{figure}[ht]
\begin{center}
\includegraphics[width=0.9\textwidth]{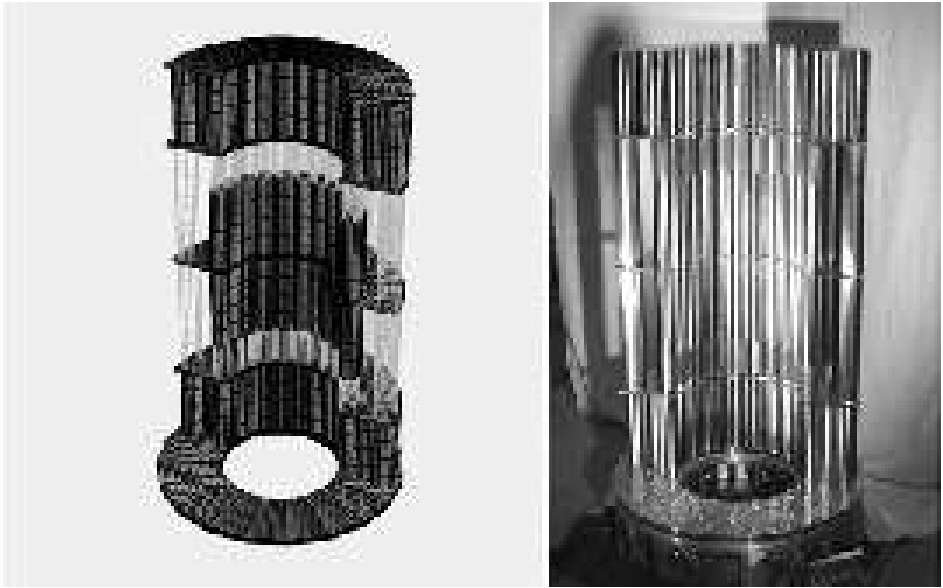}
\caption{Rendition of the ISLV source mass on left and realization on right. 
\label{islvsource}}
\end{center}
\end{figure}

The ISLV source mass is designed in a fashion similar to the pendulum and was assembled at U. Washington.  All {\it m} = 1 Newtonian derivatives of the potential below {\it l} = 8 are eliminated.  The even {\it l}, {\it m} = 1 potentials are naturally zero because of source-mass inversion symmetry about the horizontal mid-plane.  Our design therefore has four degrees of freedom used to null the 331, 551, 771, and 220 potentials.  The 220 potential is eliminated to render the source mass insensitive to small tilts about a horizontal axis, which would generate a 221 moment.  The resulting design configuration is shown in Figure \ref{islvsource}.  The mass is assembled by stacking several hundred precision-machined solid stainless steel cylinders with a mass totaling about 1500 kg.

Several trim masses are located at the outer radius of the stack in order to remove the residual 221 and 331 field moments resulting from design deviations in the fabrication of the cylinders or their stacking.  If necessary, the 441 field moment can also be trimmed.  The measurement of these residual fields is explained below.

The segmentation of the source mass into these many small cylinders allows us to measure and correct for small mass differences resulting from machining tolerances and density variations.  The ends of the cylinders have shallow male/female mating interfaces.  As each cylinder must be placed to within 50 ${\rm \mu}$m of the design position, a 15 ${\rm \mu}$m lateral clearance in the interface allows for small corrections in the placement of each piece as the stack is assembled.  The precise placement of cylinders is monitored during assembly in real time with laser-coordinate-measuring equipment.  Hollow stainless steel tubes (shown phantom in Figure 4) provide structural spacers and support the central and top portions of the stack.

\section{Experimental Technique}

The torsional motion of the pendulum is tracked using electro-optical techniques.  The source mass, which rests upon an air bearing, is rotated through a series of static positions in azimuth.  At each position the motion of the pendulum through several torsional cycles is recorded and fit to a multi-parameter model.  An azimuthal {\it m} = 1 variation in torque on the pendulum would constitute an ISLV signal and be manifest as a corresponding variation in several of these parameters.  Commonly measured indicators of a torque would be an {\it m} = 1 variation in pendulum equilibrium orientation (deflection method) or torsional oscillation frequency (frequency method).  Each of these detection methods has advantages and drawbacks.

Prior to the more sensitive cryogenic ISLV search, we are
conducting ambient temperature measurements with a second, similar
apparatus that relies on a different scheme to detect ultra-small
torques with a torsion pendulum.  We instead measure the {\it m} =
1 variation in the amplitude of the second harmonic of the
pendulum motion as our signal (``second-harmonic
method")\citep{uwa2}.  This choice is made for two reasons. First,
the amplitude of the second harmonic is insensitive to temperature
variations of the fiber\citep{uwmoriond}.  Second, the second
harmonic amplitude is insensitive as well to tilt of the fiber
attachment point.  As mentioned earlier, these effects can be
limitations when using the deflection or frequency methods.

The temperature and tilt insensitivities of the second-harmonic method come at the possible cost of a longer integration time relative to the frequency method if the measurements are dominated by additive white noise (e.g. shot noise) rather than thermal noise of the pendulum system.  This is partly because white noise on the second harmonic amplitude scales inversely as the number of periods to the ${\rm 1/2}$ and not to the ${\rm 3/2}$ power as for the frequency method.  Additionally, measurement of a signal torque using the second-harmonic method is intrinsically nearly an order of magnitude more sensitive to white-noise timing error than when the frequency method is used, even for a single oscillation period.  At room temperature, however, using the frequency method would require the fiber temperature to be held constant to roughly a ${\rm \mu}$K to achieve the measurement goal of this experiment, a severe challenge.  However, we plan to repeat the experiment at the Hanford site using the cryogenic apparatus described earlier.  When operating at $\sim$2 K, the reduced temperature coefficient of the torsion fiber and improved temperature regulation should enable us to reap the strong advantage of shortened integration time.

The room-temperature pendulum is enclosed in a vacuum chamber evacuated to a pressure below ${\rm 10^{-6}}$ mbar where the damping of the torsion oscillations is dominated by the fiber and not by  gas in the chamber.  A magnetic damper suppresses low frequency (non-torsional) mechanical modes of the pendulum--fiber system.  A three-layer magnetic shield demonstrably renders magnetic effects negligible.

\begin{figure}[ht]
\begin{center}
\includegraphics[width=0.75\textwidth]{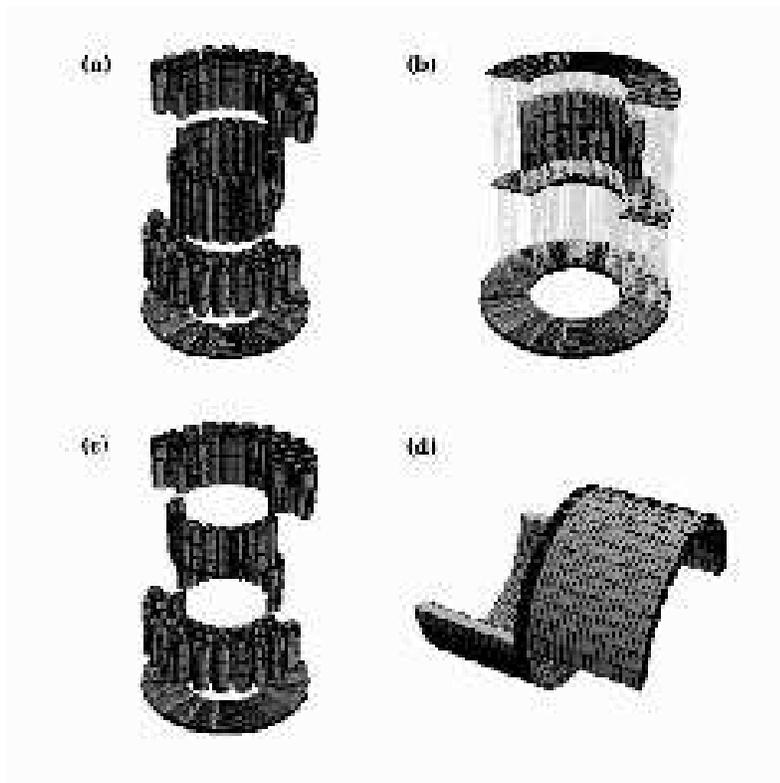}
\caption{(a) Rendition of the exaggerated 221 source mass.  (b) Rendition of the exaggerated 331 source mass.  The support structure, composed of hollow tubes (in phantom) and trays similar to those in the standard ISLV configuration, are shown as well as solid source-mass elements.  (c) Rendition of the exaggerated 441 source mass.  (d) Rendition of the exaggerated 221 pendulum.
\label{s21}\label{s31}\label{s41}\label{pend21}}
\end{center}
\end{figure}

The residual 221 and 331 mass moments of the pendulum, which result from the non-zero dimensional tolerances associated with the fabrication process, are suppressed by four movable trim masses attached to the outer face of the thicker disk shown in Figure \ref{islvpend}.  These critical mass moments are first measured by stacking the source mass in configurations that exaggerate the potentials coupling to these moments.  Figures \ref{s21}(a), \ref{s21}(b), and \ref{s21}(c) show the 221, 331, and 441 exaggerated configurations  respectively.  The 441 moment of the pendulum is not trimmed, but is measured to verify that it falls below the maximum acceptable value.

The lowest-order field moments of the source mass are also measured and if neccessary trimmed to conform to the design goals.  The pendulum used for this trimming procedure has an exaggerated 221 mass moment and is shown in Figure \ref{pend21}(d).  This pendulum is used directly to measure and trim the residual 221 field moment.  Moreover, because
\begin{equation}
V_{331} = \frac{\partial}{\partial z} V_{221}
\end{equation}
and
\begin{equation}
V_{441} = \frac{\partial^2}{\partial z^2} V_{221}
\end{equation}
the residual 331 and 441 moments can be measured with the 221 pendulum alone by measuring the 221 field at several elevations relative to the mid-plane of the source mass.  The linear and quadratic variation of the signal with elevation then yields the magnitude of the 331 and 441 fields.  Once these fields are measured, they can be trimmed away iteratively using the trim masses located at the back of the source mass stack.

\section{The Design Goals}

Uncompensated fabrication errors in the ISLV pendulum and source mass will result in spurious apparent values of the measured parameter ${\it \alpha}$ in Equation \ref{fullpot}.  To assess the resulting error in the measurement of ${\it \alpha}$ we first estimate the magnitudes of the fabrication errors based on measurements of the completed source-mass cylinders, the measured performance of our 3-D laser coordinate measuring equipment, and practical experience in manufacturing fused silica parts.  For the source mass, 50 ${\rm \mu}$m placement errors of the individual cylinders dominate their 6 ${\rm \mu}$m fabrication errors whereas for the pendulum it is the 6 ${\rm \mu}$m thickness tolerances of the fused silica components that dominate.

Departures from designs due to placement and machining errors may be correlated, so simply combining them in quadrature could underestimate their true contribution.  Instead, worst-case error estimates for mass and field moments were based on these placement and fabrication errors as maximum deviations from the three-dimensional design configuration.  Once the apparatus is fully assembled, and we are able to empirically measure the 221, 331, and 441 mass moments and field moments, the measured values should be comfortably less than these worst-case estimates.

To gauge the magnitude of the fundamental constraints posed by systematic error from residual Newtonian gravitational coupling between pendulum and source mass, Table \ref{errsig} gives the equivalent contribution to ${\it \alpha}$ (at a range ${\it \lambda}$ = 12 cm) for interactions resulting solely from the worst-case fabrication and placement errors.  We see that the limiting interactions are the 441 and 771 terms which yield a contribution to ${\it \alpha}$ of a few times ${\rm 10^{-6}}$.  Aside from these limiting systematic errors, there is also the statistical measurement uncertainty.  For practical integration times at room temperature, we may be able to constrain ${\it \alpha}$ to less than ${\rm 10^{-4}}$.  When the experiment is repeated at $\sim$2 K, the thermal noise is expected to be significantly reduced, possibly allowing an upper limit on ${\it \alpha}$ below ${\rm 10^{-5}}$.  Cryogenic operation may effectively remove noise considerations and leave the measurement limited by the Newtonian gravitational interactions.  Such a limit is implied by Table \ref{errsig}, and would yield a constraint on ${\it \alpha}$ of about 5 x ${\rm 10^{-6}}$, limited by the worst-case 441 interaction.  It may be possible to trim the 441 moment of the source mass by as much as 90\% leaving the worst-case 771 interaction as the limiting systematic effect.

\begin{table}[bt]
\caption{Gravitational second harmonic signals arising from the combined fabrication errors of the pendulum and source mass.}
{\footnotesize
\begin{center}
\begin{tabular}{@{}cc@{}}
\hline
{} &{} \\[-1.5ex]
Multipole Signal & Equivalent ${\it \alpha}$ (${\it \lambda}$ = 12 cm) \\[1ex]
\hline
{} &{} \\[-1.5ex]
221 & 8.22 ${\times 10^{-7}}$ \\[1ex]
331 & 1.32 ${\times 10^{-7}}$ \\[1ex]
441 & 4.60 ${\times 10^{-6}}$ \\[1ex]
551 & 3.58 ${\times 10^{-7}}$ \\[1ex]
771 & 1.31 ${\times 10^{-6}}$ \\[1ex]
991 & 3.23 ${\times 10^{-8}}$ \\[1ex]
\hline
\end{tabular}\label{errsig}
\end{center}
}
\end{table}

\begin{figure}[tripod]
\begin{center}
\includegraphics[width=0.5\textwidth]{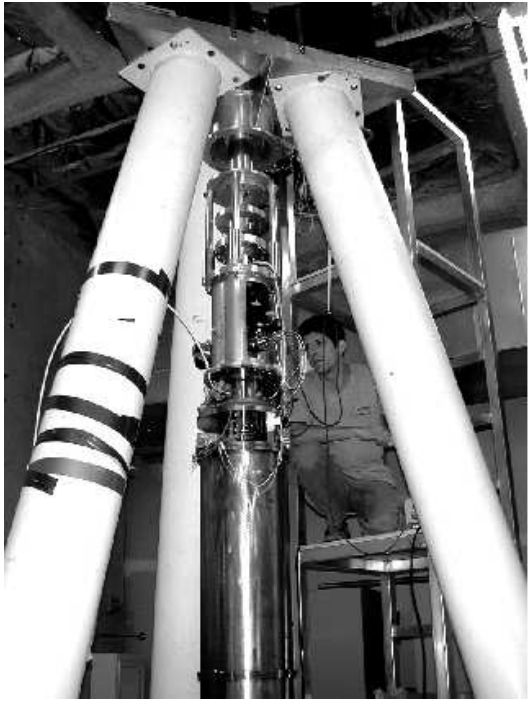}
\caption{Graduate student Ricco Bonicalzi adjusting room-temperature torsion pendulum apparatus in U. Washington lab (magnetic shields absent).
\label{tripod}}
\end{center}
\end{figure}

\begin{figure}[optics]
\begin{center}
\includegraphics[width=0.5\textwidth]{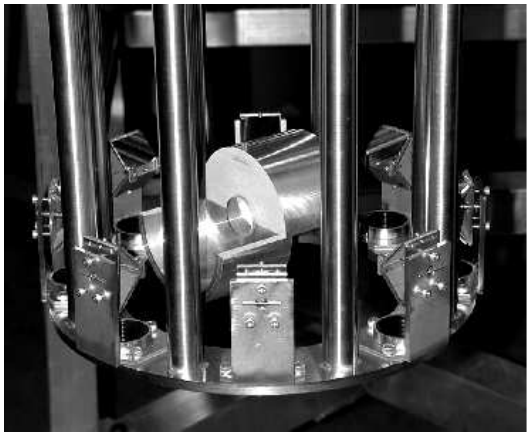}
\caption{Photo showing 221 pendulum (see Figure 5(d)) with vacuum can removed.  The pendulum is surrounded by mirrors that direct collimated laser beams to and away from mirror on pendulum.  As pendulum rotates, each departing beam sweeps across a split photodiode and the time of passage through a fiducial point is measured.  These "crossing times" are used to determine details of pendulum motion, and ultimately the magnitude of any torque acting on the pendulum.
\label{optics}}
\end{center}
\end{figure}

\begin{figure}[cryostat]
\begin{center}
\includegraphics[width=0.5\textwidth]{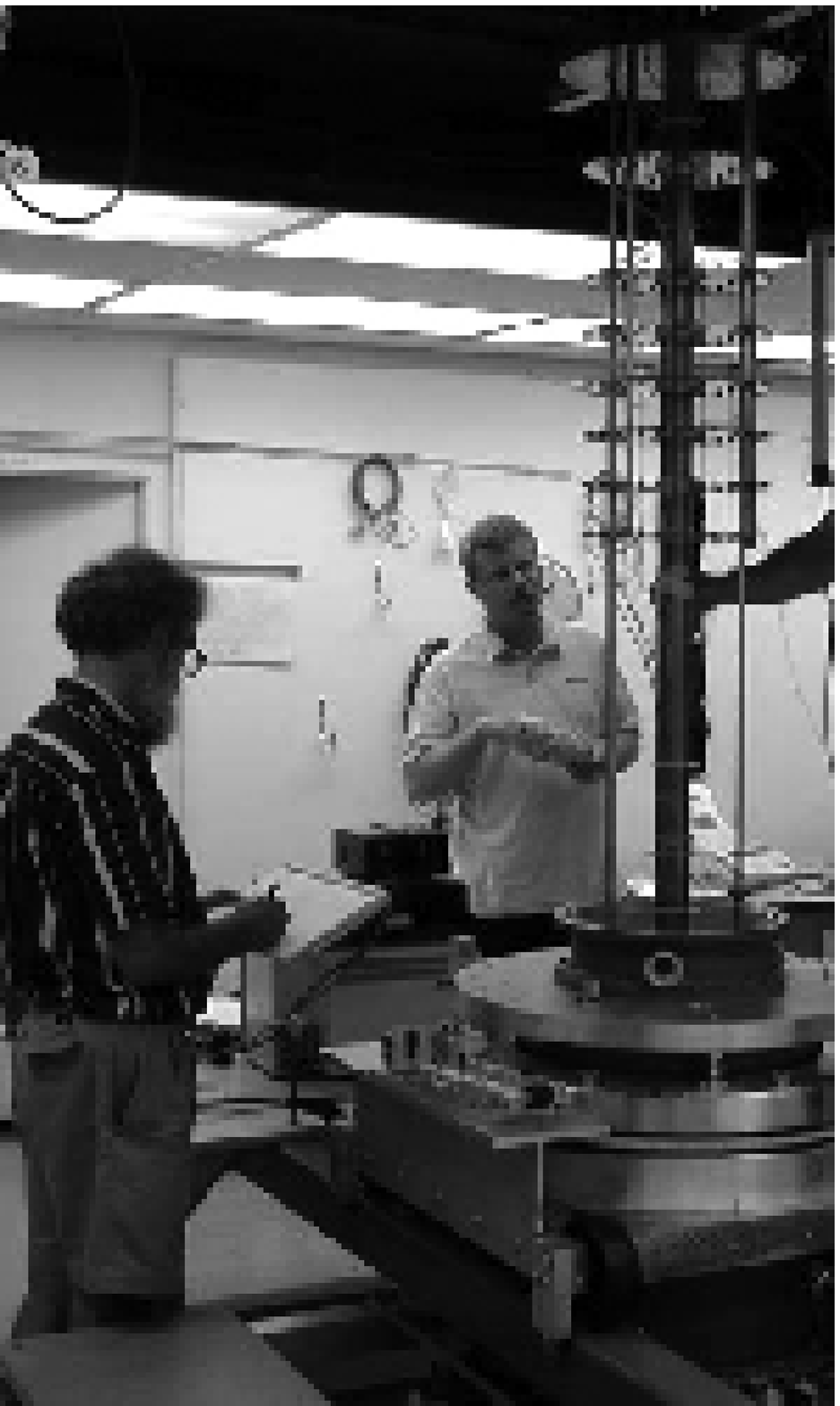}
\caption{Cryostat (far right) being withdrawn from dewar at BGPL.
Riley Newman (left) and Michael Moore observe the process.
\label{cryostat}}
\end{center}
\end{figure}

The room-temperature apparatus shown in Figures, \ref{tripod} and
\ref{optics} has been assembled and tested at the University of
Washington and is being moved to the BGPL in July 2006 to be
reassembled.  We expect much improved performance in this
ultra-quiet site compared to the semi-urban vibration environment
of our campus laboratory.  The Boeing laser metrology equipment
will also be relocated to enable reassembly of the source mass.
This torsion pendulum instrument will be used first to trim and
verify the field moments of the ISLV source mass by direct
gravitational measurements.  As described earlier, the exaggerated
source mass configurations will allow us to trim the pendulum mass
moments to meet the design requirements also.  At this point we
will attempt an ISLV search with this lower-sensitivity apparatus,
followed by installation of the ISLV pendulum in the cryostat
shown in Figure \ref{cryostat} in order to search down to ${\it
\alpha}$ around ${\rm 10^{-5}}$ or below.  We also intend to use
the cryogenic apparatus for a high-sensitivity equivalence
principal test\citep{ep}.

\section*{Acknowledgments}
We would like to thank and to acknowledge the support of the
Pacific Northwest National Laboratory, our experimental site host;
and to the Boeing Company for the long-term loan of
state-of-the-art laser coordinate measuring equipment.  This
research is funded under National Science Foundation grants
PHY-0108937, PHY-9803765 and PHY-0244762.

Paul Boynton offers his contribution to these proceedings as a
token tribute to the memory of Francesco Melchiorri, whom he had
the great privilege to know as a true friend for many years.

\end{document}